\documentstyle[prl,aps]{revtex}

\begin{document}
\author{Jian-Qi Shen $^{1,}$$^{2}$ \footnote{E-mail address: jqshen@coer.zju.edu.cn}}
\address{$^{1}$  Centre for Optical
and Electromagnetic Research, State Key Laboratory of Modern
Optical Instrumentation, \\Optical Engineering Department,
Zhejiang University,
Hangzhou Yuquan 310027, P.R. China\\
$^{2}$ Zhejiang Institute of Modern Physics and Department of
Physics, Zhejiang University, Hangzhou 310027, P.R. China}
\date{\today }
\title{The connection between conformal group and quantum states of photons\footnote{This note is mainly devoted
to a physically interesting comparison between the photonic
quantum states and the conformal group. Because of the trivial and
lengthy calculation involved, the detailed analysis of comparison
made between photonic states and conformal transformations will be
submitted nowhere else for publication, just uploaded at the
e-print archives. }} \maketitle

\begin{abstract}
This note is concerned with the connections between the conformal
group and the quantum states of photons. It is shown that there
exist analogies between the photonic quantum states and the
conformal group, namely, the time-development operator (with a
free Hamiltonian), displacement and squeezing operators of the
vacuum state corresponds to the dilatation, translation, proper
Lorentz transformations, respectively, and that the three quantum
states of photons ({\it i.e.}, Fock state, coherent state and
squeezed state) in quantum optics thus bear some analogy to the
above three transformations in the conformal group. Based on this
comparison, we argue by analogy that if the phase transformation
operators acting on a vacuum state (hence the Fock state, coherent
state and squeezed state are generated) are truly exactly
analogous to the conformal transformations, then a fourth quantum
state of photons (referred to as the {\it conformal state}), which
corresponds to the {\it special conformal} transformation
(acceleration transformation) and will therefore be of special
physical interest, can be suggested.
\end{abstract}

\section{The 15-parameter conformal group}
The group under consideration is the 15-parameter Lie group often
referred to as ``conformal transformation''\cite{Fulton,Flato},
which is defined as the set of those transformations that
transforms flat space into flat space. The conformal group
consists of the space-time translations
($x'^{\mu}=x^{\mu}+\alpha^{\mu}$, 4 parameters), the proper
homogeneous Lorentz transformations ({\it i.e.}, the Lorentz
rotation, $x'^{\mu}=\Lambda^{\mu}_{\nu}x^{\nu}$, 6 parameters),
the dilatation (or scale) transformation ($x'^{\mu}=sx^{\mu}$, 1
parameter) and the special conformal (acceleration) transformation
($x'^{\mu}=(1+2a^{\alpha}x_{\alpha}+x^{2}a^{2})^{-1}(x^{\mu}+a^{\mu}x^{2})$,
4 parameters). The corresponding algebraic generators can be
realized in terms of the differential operators acting on the
Minkowski space, which are as follows\cite{Fulton}:
\begin{eqnarray}
P_{\mu}&=&i\partial_{\mu}    \quad  ({\rm translation}),                 \nonumber \\
M_{\mu\nu}&=&i(x_{\mu}\partial_{\nu}-x_{\nu}\partial_{\mu}) \quad
({\rm Lorentz   \quad  transformation}),
\nonumber \\
D&=&ix^{\nu}\partial_{\nu}              \quad     ({\rm dilatation
  \quad  transformation}),
\nonumber \\
 K_{\mu}&=&i(2x_{\mu}x_{\nu}\partial^{\nu}-x^{2}\partial_{\mu})      \quad     ({\rm special
 \quad  conformal   \quad   transformation}).
\end{eqnarray}

The algebraic commuting relations of the above 15 generators are
given\cite{Fulton}

\begin{eqnarray}
     [P_{\mu}, P_{\nu}]&=&0,  \quad  [P_{\lambda}, M_{\mu\nu}]=i(g_{\mu\lambda}P_{\nu}-g_{\nu\lambda}P_{\mu}),             \nonumber \\
       \left[P_{\mu}, D\right]&=&iP_{\mu},    \quad    [P_{\mu},
       K_{\nu}]=2i(g_{\mu\nu}D-M_{\mu\nu}),
                  \nonumber \\
               \left[D, M_{\mu\nu}\right]&=&0,         \quad   [D,
               K_{\mu}]=iK_{\mu},
                  \nonumber \\
             \left[M_{\mu\nu},
             M_{\sigma\rho}\right]&=&i(g_{\mu\rho}M_{\nu\sigma}+g_{\nu\sigma}M_{\mu\rho}+g_{\mu\sigma}M_{\rho\nu}+g_{\nu\rho}M_{\sigma\mu}),
                  \nonumber \\
                  \left[K_{\mu}, K_{\nu}\right]&=&0,   \quad    [K_{\lambda},
                  M_{\mu\nu}]=i(g_{\mu\lambda}K_{\nu}-g_{\nu\lambda}K_{\mu}).
\end{eqnarray}
In what follows we will concern ourselves with the unitary phase
transformation operators in the photonic quantum states, {\it
i.e.}, the Fock state, coherent state and squeezed state.

\section{Photonic quantum states}
Historically, the fundamental concepts of the Fock state, coherent
state and squeezed state were proposed by Dirac, Glauber and
Stoler\cite{Dirac,Glauber,Stoler}, respectively. In this section
we will take into a comprehensive consideration these three
photonic quantum states and discuss several topics such as the
unitary phase transformation operators, the generators and the
algebraic structures in quantum states of photons as well as the
close relation to the conformal transformation\cite{Li}.

\subsection{Free-Hamiltonian time-evolution operator in Fock state}
The operators $a^{\dagger}a$ and $aa^{\dagger}$ may be considered
the generators of the Free-Hamiltonian time-evolution operator of
photon state, which is of the form
\begin{equation}
P(\alpha)=\exp (\alpha a^{\dagger}a-\alpha^{\ast}aa^{\dagger}),
\quad                        P^{\dagger}(\alpha)=\exp [-(\alpha
a^{\dagger}a-\alpha^{\ast}aa^{\dagger})].    \label{a1}
\end{equation}
The unitary transformation operator $P(\alpha)$ leads to the
following transformation
\begin{equation}
a\rightarrow
a'=P^{\dagger}(\alpha)aP(\alpha)=\exp(\alpha^{\ast}-\alpha)a,
\quad
a^{\dagger}\rightarrow
{a'^{\dagger}}=P^{\dagger}(\alpha)a^{\dagger}P(\alpha)=\exp[-(\alpha^{\ast}-\alpha)]a^{\dagger},
\end{equation}
and consequently the infinitesimal variations of $a$ and
$a^{\dagger}$ under the transformation $P(\alpha)$ are
\begin{equation}
\delta a=(\alpha^{\ast}-\alpha)a,  \quad    \delta
a^{\dagger}=-(\alpha^{\ast}-\alpha)a^{\dagger},
\end{equation}
which follows that $\delta a$, $a^{\dagger}$ and the infinitesimal
variations of space-time coordinates
\begin{equation}
\delta x^{\mu}=\frac{1}{i}\alpha Dx^{\mu}=\alpha x^{\mu}
\end{equation}
under the dilatation transformation are alike in some way. The new
states generated by the dilatation transformation $P(\alpha)$ are
written
\begin{equation}
|\alpha, 0\rangle=P(\alpha)|0\rangle=\exp
(\alpha^{\ast})|0\rangle, \quad           |\alpha,
n\rangle=P(\alpha)|n\rangle=\exp
[(\alpha^{\ast}-\alpha)a^{\dagger}a]\exp (\alpha^{\ast})|n\rangle.
\label{eq7}
\end{equation}
It follows that $|\alpha, n\rangle=P(\alpha)|n\rangle=\exp
[n(\alpha^{\ast}-\alpha)]\exp (\alpha^{\ast})|n\rangle$. It is
known that the quantum state $|\alpha, n\rangle$ can be realized
by the time evolution with a free Hamiltonian
$H=\frac{\omega}{2}(aa^{\dagger}+a^{\dagger}a)$ (governed by the
time-dependent Schr\"{o}dinger equation), namely, $|t,
n\rangle=\exp [\frac{1}{i}(n+\frac{1}{2})\omega t]|n\rangle$. If
the parameters in the unitary transformation operator $P(\alpha)$
are taken
\begin{equation}
\alpha=-\frac{1}{2i}\omega t,     \quad
\alpha^{\ast}=\frac{1}{2i}\omega t,
\end{equation}
then $|t, n\rangle$ will be the photonic states characterized in
(\ref{eq7}). Thus it is concluded that the free-Hamiltonian
time-evolution operator acting on the photon creation and
annihilation operators closely resembles the dilatation (or scale)
transformation in the conformal group, and that a stationary Fock
state acted upon by the free-Hamiltonian time-evolution operator
$P(\alpha)$ will be transformed into a time-dependent Fock state.

\subsection{Displacement operator in coherent state}
The coherent state of photons is defined to be
$|\alpha\rangle=D(\alpha)|0\rangle$ with the displacement operator
being
\begin{equation}
D(\alpha)=\exp(\alpha a^{\dagger}-\alpha^{\ast}a).
\end{equation}
The displacement operator acting on the vacuum state is equivalent
to the following transformation
\begin{equation}
a\rightarrow a'=D^{\dagger}(\alpha)aD(\alpha)=a+\alpha, \quad
a^{\dagger}\rightarrow{a'^{\dagger}}=D^{\dagger}(\alpha)a^{\dagger}D(\alpha)=a^{\dagger}+\alpha^{\ast},
\end{equation}
which will yield the infinitesimal variations
\begin{equation}
\delta a=\alpha,        \quad    \delta a^{\dagger}=\alpha^{\ast}.
\end{equation}
It is physically interesting that the above $\delta a$ and $\delta
a^{\dagger}$ are in analogy with the variations of the space-time
coordinates $x^{\mu}$ under the infinitesimal translation
transformation in the conformal group, {\it i.e.},
\begin{equation}
\delta x^{\mu}=\frac{1}{i}
\alpha^{\nu}P_{\nu}x^{\mu}=\alpha^{\mu}.
\end{equation}

\subsection{Squeezing operator in squeezed state}
The squeezed state is defined to be
$|\zeta\rangle=S(\zeta)|0\rangle$ with the squeezing operator
being
\begin{equation}
S(\zeta)=\exp\left[\frac{1}{2}\zeta^{\ast}a^{2}-\frac{1}{2}\zeta(a^{\dagger})^{2}\right],
\quad \zeta=s\exp(i\theta).
\end{equation}
The corresponding variations of $a$ and $a^{\dagger}$ are
\begin{equation}
a\rightarrow a'=S^{\dagger}(\alpha)aS(\alpha)=a\cosh
s-a^{\dagger}\exp(i\theta)\sinh s,                \quad
a^{\dagger}\rightarrow{a'^{\dagger}}=S^{\dagger}(\alpha)a^{\dagger}S(\alpha)=a^{\dagger}\cosh
s-a\exp(-i\theta)\sinh s.
\end{equation}
The infinitesimal variations
\begin{equation}
\delta
a=\left[-\left(\frac{1}{2}\zeta^{\ast}a^{2}-\frac{1}{2}\zeta(a^{\dagger})^{2}\right),
a\right]=-\zeta a^{\dagger},      \quad      \delta
a^{\dagger}=\left[-\left(\frac{1}{2}\zeta^{\ast}a^{2}-\frac{1}{2}\zeta(a^{\dagger})^{2}\right),
a^{\dagger}\right]=-\zeta^{\ast} a,
\end{equation}
which shows some analogy with the two-dimensional infinitesimal
Lorentz rotation, $\delta
x^{\mu}=\frac{1}{2i}\epsilon^{\omega\nu}M_{\omega\nu}x^{\mu}=\epsilon^{\nu\mu}x_{\nu}$,
{\it i.e.},
\begin{equation}
\delta x^{0}=-\epsilon^{10}x^{1},    \quad                 \delta
x^{1}=-\epsilon^{10}x^{0}.
\end{equation}
\section{Defining a kind of operator integral to obtain the generators of quantum states of photons}
In this section, we will define a kind of operator integral
technique to get the algebraic generators of the above three
quantum states of photons.

The generators of displacement operator $D(\alpha)=\exp(\alpha
a^{\dagger}-\alpha^{\ast}a)$ can be obtained via the following two
integrals (with c-number $\alpha$ and $\alpha^{\ast}$ being the
integrands)
\begin{equation}
{\alpha^{\ast}}a=\int {\alpha^{\ast}{\rm d}a},    \quad {\alpha}
a^{\dagger}=\int { \alpha {\rm d}a^{\dagger}}, \label{eq17}
\end{equation}
where the integral constant (unit matrix) is omitted due to its
triviality. In (\ref{eq17}) we obtain the linear-form operators
(generators) $a$ and $a^{\dagger}$.

In the similar fashion, we calculate the following operator
integrals (note that the definition of the operator integral is
implied in the following calculation)
\begin{eqnarray}
{\mathcal F}&=&\int {\rm d}a\left(\zeta
a^{\dagger}+\zeta^{\ast}a\right)=\frac{1}{2}\zeta
\left(aa^{\dagger}+a^{\dagger}a\right)+\frac{1}{2}\zeta^{\ast}a^{2},                  \nonumber \\
{\mathcal F}^{\dagger}&=&\int {\rm d}a^{\dagger}\left(\zeta
a^{\dagger}+\zeta^{\ast}a\right)=\frac{1}{2}\zeta^{\ast}
\left(aa^{\dagger}+a^{\dagger}a\right)+\frac{1}{2}\zeta
\left(a^{\dagger}\right)^{2}.
\end{eqnarray}
So, one can arrive at the generators of squeezed state (and Fock
state), which are involved in
\begin{equation}
{\mathcal F}-{\mathcal
F}^{\dagger}=\frac{1}{2}\left[\zeta^{\ast}a^{2}-\zeta
\left(a^{\dagger}\right)^{2}\right]+\frac{1}{2}\left(\zeta-\zeta^{\ast}\right)\left(aa^{\dagger}+a^{\dagger}a\right).
\label{eq19}
\end{equation}
Note that the generators in (\ref{eq19}) are quadratic-form
operators. In Eq.(\ref{eq19}) $aa^{\dagger}$ and $a^{\dagger}a$
can be viewed as the generators of the dilatation transformation
$P(\alpha)$ in (\ref{a1}). If, for example,
$\frac{1}{2i}\left(\zeta-\zeta^{\ast}\right)={\rm Im}\alpha$ and
${\rm Re}\alpha=0$, then we have
$\frac{1}{2}\left(\zeta-\zeta^{\ast}\right)\left(aa^{\dagger}+a^{\dagger}a\right)=\alpha
a^{\dagger}a-\alpha^{\ast}aa^{\dagger}$.

The calculations in Eq.(\ref{eq17}) and (\ref{eq19}) shows that
one can obtain the generators of photonic quantum states by using
such operator integrals just defined above.
\\ \\

Now we continue calculating the following operator integrals
\begin{eqnarray}
{\mathcal G}&=&\int {\rm d}a\left[\varrho^{\ast}a^{2}+\varrho\left(a^{\dagger}\right)^{2}\right]=\frac{\varrho}{3} \left[a\left(a^{\dagger}\right)^{2}+a^{\dagger}aa^{\dagger}+(a^{\dagger})^{2}a\right]+\frac{\varrho^{\ast}}{3}a^{3}=\varrho a^{\dagger}aa^{\dagger}+\frac{\varrho^{\ast}}{3}a^{3},                    \nonumber \\
{\mathcal G}^{\dagger}&=&\int {\rm
d}a^{\dagger}\left[\varrho^{\ast}a^{2}+\varrho\left(a^{\dagger}\right)^{2}\right]=\frac{\varrho^{\ast}}{3}
\left[\left(a^{\dagger}\right)a^{2}+aa^{\dagger}a+a^{2}a^{\dagger}\right]+\frac{\varrho}{3}\left(a^{\dagger}\right)^{3}=\varrho^{\ast}aa^{\dagger}a+\frac{\varrho}{3}\left(a^{\dagger}\right)^{3}.
\end{eqnarray}
In the same manner as (\ref{eq19}), we have
\begin{equation}
{\mathcal G}-{\mathcal G}^{\dagger}=\varrho Q-\varrho^{\ast}
Q^{\dagger},
\end{equation}
where
\begin{equation}
Q=a^{\dagger}aa^{\dagger}-\frac{1}{3}\left(a^{\dagger}\right)^{3},
\quad Q^{\dagger}=aa^{\dagger}a-\frac{1}{3}a^{3},
\end{equation}
which are nonlinear (cubic-form) generators.

\section{Cubic-nonlinearity phase transformation operator}
It is of physical interest to consider the so-called
cubic-nonlinearity phase transformation operator $C(\varrho)$,
which is defined to be
\begin{equation}
C(\varrho)=\exp \left(\varrho Q-\varrho^{\ast} Q^{\dagger}\right).
\end{equation}
In the meanwhile, we define a new photonic quantum state as
follows
\begin{equation}
|\varrho\rangle=C(\varrho)|0\rangle.
\end{equation}
The variations of $a$ and $a^{\dagger}$ under the nonlinear
unitary transformation $C(\varrho)$ are
\begin{eqnarray}
\delta a&=&\left[-\left(\varrho Q-\varrho^{\ast}
Q^{\dagger}\right),
a\right]=\varrho\left(aa^{\dagger}+a^{\dagger}a\right)-\left[\varrho^{\ast}
a^{2}+\varrho\left(a^{\dagger}\right)^{2}\right],     \nonumber   \\
\delta a^{\dagger}&=&\left[-\left(\varrho Q-\varrho^{\ast}
Q^{\dagger}\right),
a^{\dagger}\right]=\varrho^{\ast}\left(aa^{\dagger}+a^{\dagger}a\right)-\left[\varrho\left(a^{\dagger}\right)^{2}+\varrho^{\ast}
a^{2}\right].    \label{eq25}
\end{eqnarray}
Accordingly, here we may take into account the special conformal
transformation in the conformal group, {\it i.e.},
\begin{equation}
\delta
x^{\mu}=\frac{1}{i}\varrho^{\nu}K_{\nu}x^{\mu}=2\varrho^{\nu}x_{\nu}x^{\mu}-x^{2}\varrho^{\mu},
\end{equation}
that is, the two-dimensional infinitesimal special conformal
transformation is of the form
\begin{eqnarray}
\delta
x^{0}&=&2\varrho^{\nu}x_{\nu}x^{0}-x^{2}\varrho^{0}=-\left\{{2\varrho^{1}x^{1}x^{0}-\varrho^{0}\left[\left(x^{0}\right)^{2}+\left(x^{1}\right)^{2}\right]}\right\},
\nonumber  \\
x^{1}&=&2\varrho^{\nu}x_{\nu}x^{1}-x^{2}\varrho^{1}=2\varrho^{0}x^{0}x^{1}-\varrho^{1}\left[\left(x^{0}\right)^{2}+\left(x^{1}\right)^{2}\right].
\label{eq27}
\end{eqnarray}
It follows that the quantum state $C(\varrho)|0\rangle$ and the
special conformal transformation (characterized by the generators
$K_{\mu}$'s) are alike in some way. If, for example, in the
particular case the two-dimensional infinitesimal parameters agree
with $\varrho^{0}=-\varrho^{1}$, and $\varrho=\varrho^{\ast}$,
then the two variations (\ref{eq25}) and (\ref{eq27}) are of the
same mathematical form. Both (\ref{eq25}) and (\ref{eq27}) are
their respective extensions of this same special case. Thus we may
think of $|\varrho\rangle=C(\varrho)|0\rangle$ as the fourth
quantum state of photons and refer to it as the optical {\it
conformal state}. Differing from the three photonic quantum states
studied previously, the conformal state is a nonlinear one. It is
reasonably believed that such quantum state may be of special
physical interest and deserves further investigation.
\\ \\

\textbf{Acknowledgements}  This project was supported partially by
the National Natural Science Foundation of China under the project
No. $90101024$.

\end{document}